\documentclass{article}
\begin{document}
\newcommand{\be}{\begin{eqnarray}}
\newcommand{\ee}{\end{eqnarray}}
\newcommand{\ra}{\rightarrow}
\newcommand{\la}{\leftarrow}
\newcommand{\noi}{\noindent}
\newcommand{\rang}{\rangle}
\newcommand{\lang}{\langle}
\title{Local Hidden Variable Theoretic Measure of
Quantumness of Mutual Information
\footnote{J.Phys. A:Math.Theor.{\bf 47}, 115303 (2014)}}
\author{R.R.Puri\footnote{e-mail: ravirpuri@gmail.com}\\
Department of Physics, Indian Institute of Technology-Bombay,\\
Powai, Mumbai-400076.}
\date{}
\maketitle
\noi

\section{Introduction}~\label{sec1}

Entanglement, a manifestation of quantumness of correlations between
observables of the subsystems of a composite system, once thought to
be an essential ingredient for distinctive quantum features in
quantum information processing, is no longer considered to be
so as it is found that the unique features of quantum information 
processing are contained in the
quantum nature of mutual information which does not necessarily require
entanglement (see~\cite{modi} and references therein). Whereas 
the concept of quantumness of correlations between observables of the
parts of a composite system is characterized by their 
incommensurability with the predictions of the local hidden variable 
(LHV) theory (see~\cite{horo2} and references therein), that of the
quantumness of information does not invoke the LHV theory
~\cite{olivier}-\cite{wu} explicitly. Different protocols for identifying 
classical content of information lead to different measures of 
quantumness of mutual information like quantum 
discord~\cite{olivier}, quantum deficit~\cite{horo1}, measurement
induced disturbance~\cite{luo1}, symmetric discord~\cite{wu} and
others~\cite{modi}. A number of analytic and numerical results for
these measures of quantumness for various states of two qubits have
been reported~\cite{olivier}-\cite{sarandy}. These results show that
even a separable state may contain quantum features in its information
content.

In this paper a measure of  quantumness of mutual information is 
proposed by invoking the LHV theory  explicitly. The proposed measure 
turns out to be useful as it circumvents the need of optimization of 
classical information over possible directions of measurement for a 
class of states and simplifies finding optimized classical information 
for others. Moreover,  under specific situations, it fits in with one 
or the other widely used measures, namely, the measurement 
induced disturbance, the symmetric discord, and the quantum discord.

To that end, the classical mutual information $I_{{\rm LHV}}$ in this
paper is defined following the LHV theoretic considerartions
of~\cite{rrp} regarding characterization of quantumness of
correlations between observables in a system of spins A and B
described by the density operator $\hat\rho^{\rm AB}$. 
It is used to measure the quantumness of mutual information as 
$Q_{{\rm LHV}}=I_Q(\hat\rho^{\rm AB})-I_{\rm LHV}$, where 
$I_Q(\hat\rho^{\rm AB})$ is the quantum information in 
$\hat\rho^{\rm AB}$. The $Q_{{\rm LHV}}$ is found to be
identical with the  measurement induced disturbance if the Bloch 
vectors $\langle\hat{\bf S}^{\rm A}\rangle$ 
and $\langle\hat{\bf S}^{\rm B}\rangle$ of spins A and B
are non-zero where $\hat{\bf S}^{\rm A}$ ($\hat{\bf S}^{\rm B}$) is 
the spin vector of spin A (spin B) and $\langle\hat P\rangle
={\rm Tr}(\hat P\hat\rho^{{\rm AB}})$. If 
$\langle\hat{\bf S}^{\rm A}\rangle=
\langle\hat{\bf S}^{\rm B}\rangle=0$ then $I_{{\rm LHV}}$ is the
maximimum value of classical mutual information over directions of 
measurement of the two spins which can be evaluated analytically 
exactly. The $Q_{{\rm LHV}}$ then turns out to  be the same as the 
symmetric discord. If one of the Bloch vectors, say, 
$\langle\hat{\bf S}^{\rm A}\rangle=0$, but the other
is not then, for certain states, $Q_{{\rm LHV}}$ is same as the 
quantum discord for measurement over A. Thus the
LHV theoretic quantumness of mutual information and the measurement 
induced disturbance are identical when the Bloch vector of each spin is 
non-zero. However, whereas the mesurement induced disturbance is 
non-unique when the Bloch vector of either or both the spins is zero,
the LHV theoretic measure determines the quantumness of imutual 
information uniquely even in those situations.

The paper is organized as follows. The section \ref{sec2} presents
the formulation of LHV theoretic quantumness of mutual information. It 
is compared with other measures in section \ref{sec3}. The conclusions
are summarized in section \ref{sec4}.

\section{Local Hidden Variable Theory and Quantumness}~\label{sec2}

Let us recall that a spin-1/2 in LHV theory is regarded as
a vector ${\bf S}$ in the real three dimensional space whose component 
along any direction can assume two
values, say $\pm 1/2$, and is assumed to be under the influence of
some unknown {\it hidden} causes or variables acting randomly. The random
influence of the hidden variables results in the components of the spin
in any direction acquiring randomly the values $\pm 1/2$. The
properties of the spin may then be described in terms of the
probability distribution functions $f(S_{a_1},S_{a_2},
\ldots,S_{a_N})$ for the components of the spin in the directions
${\bf a}_1,{\bf a}_2,\ldots,{\bf a}_N$
where
\be
S_{a_i}={\bf S}\cdot{\bf a}_i,\qquad |{\bf a}_i|=1,
\label{new2}
\ee
is the component of spin in the direction ${\bf a}_i$.
Now, let $p(\epsilon_{a_1},\epsilon_{a_2},\ldots,\epsilon_{a_N})$
($\epsilon_{a_i}=\pm 1$) denote the joint probability for the
components of the spin along the directions ${\bf a}_1,{\bf a}_2,
\ldots,{\bf a}_N$ to have the values $\epsilon_{a_1}/2,\epsilon_{a_2}/2,
\ldots,\epsilon_{a_N}/2$ respectively so that
\be
&&f\left(S_{a_1},S_{a_2},\ldots,S_{a_N}\right)\nonumber\\
&&=\sum_{\epsilon_{a_1},\epsilon_{a_2},
\ldots,\epsilon_{a_N}=\pm 1}\left[\delta\left(S_{a_1}-
\frac{\epsilon_{a_1}}{2}\right)
\delta\left(S_{a_2}-\frac{\epsilon_{a_2}}{2}\right)
\cdots\delta\left(S_{a_N}-\frac{\epsilon_{a_N}}{2}\right)\right]
\nonumber\\
&&~~~~~~~~~~~~~~~~~~~~
\times p(\epsilon_{a_1},\epsilon_{a_2},\ldots,\epsilon_{a_N}).
\label{new2m}
\ee
It is straightforward to invert this relation to get
\be
p(\epsilon_{a_1},\epsilon_{a_2},\ldots,\epsilon_{a_N})
=\Big\langle\Big(\frac{1}{2}+\epsilon_{a_1}S_{a_1}\Big)
\Big(\frac{1}{2}+\epsilon_{a_2}S_{a_2}\Big)
\cdots\Big(\frac{1}{2}+\epsilon_{a_N}S_{a_N}\Big)\Big\rangle,
\label{new1}
\ee
where the angular bracket denotes average with respect to the
distribution function $f\left(S_{a_1},S_{a_2},\ldots,S_{a_N}\right)$:
\be
\langle P\rangle=\int~P f\left(S_{a_1},S_{a_2},\ldots,S_{a_N}\right)
\prod_{i=1}^{N}{\rm d}S_{a_i}.
\label{new2p}
\ee
The joint probability distribution for two spins can be similarly
defined and shown to be expressible as
\be
&&p(\epsilon^{A}_{a_1},\epsilon^{A}_{a_2},\ldots,
\epsilon^{A}_{a_M};\epsilon^{B}_{b_1},\epsilon^{B}_{b_2},
\ldots,\epsilon^{B}_{b_N})\nonumber\\
&&=\Big<\Big(\frac{1}{2}+\epsilon^{A}_{a_1}S_{a_1}\Big)
\Big(\frac{1}{2}+\epsilon^{A}_{a_2}S_{a_2}\Big)
\cdots\Big(\frac{1}{2}+\epsilon^{A}_{a_M}S_{a_M}\Big)
\nonumber\\
&&~~~~~~\Big(\frac{1}{2}+\epsilon^{B}_{b_1}S_{b_1}\Big)
\Big(\frac{1}{2}+\epsilon^{B}_{b_2}S_{b_2}\Big)
\cdots\Big(\frac{1}{2}+\epsilon^{B}_{b_N}S_{b_N}\Big)
\Big>,
\label{new3}
\ee
where $p(\epsilon^{A}_{a_1},\epsilon^{A}_{a_2},\ldots,
\epsilon^{A}_{a_M};\epsilon^{B}_{b_1},\epsilon^{B}_{b_2},
\ldots,\epsilon^{B}_{b_N})$ stands for the probabilty of finding
the components of spin A to have values
$\epsilon^{A}_{a_1}/2,\epsilon^{A}_{a_2}/2,\ldots,
\epsilon^{A}_{a_M}/2$
in the directions ${\bf a}_1,{\bf a}_2,\ldots,{\bf a}_M$, and
the components of spin B to have the values
$\epsilon^{B}_{b_1}/2$,$\epsilon^{B}_{b_2}/2,\ldots,$
$\epsilon^{B}_{b_N}/2$ in the directions
${\bf b}_1,{\bf b}_2,\ldots,{\bf b}_N$ with
$\epsilon^{A}_{a_i},\epsilon^{B}_{b_i}=\pm 1$.
The form (\ref{new3}) for the joint probability is useful for 
constructing its quantum analog by (i) replacing the
classical random variables $S_{a}$ by the operators $\hat S_{a}$
which obey the commutation relation
\be
[\hat S_{a},~\hat S_{b}]={\rm i}({\bf a}
\times{\bf b})\cdot\hat{\bf S},
\label{comel}
\ee
where $\hat{\bf S}={\bf e}_1\hat S_{e_1}+{\bf e}_2\hat S_{e_2}
+{\bf e}_3\hat S_{e_3}$ (${\bf e}_i\cdot{\bf e}_j=\delta_{ij}$),
and the anti-commutation relation
\be
\hat S_a\hat S_b+\hat S_b\hat S_a=\frac{{\bf a}\cdot{\bf b}}{2},
\label{new9}
\ee
(ii) by assigning a rule, called the Chosen Ordering, for
ordering operators in a product of non-commuting operators, and
(iii) by replacing the average therein as the quantum mechanical
expectation value wherein the system is described by a density matrix
$\hat\rho$ and the expectation value of an operator $\hat P$ is given
by $\lang\hat P\rang={\rm Tr}(\hat P\hat\rho)$. This approach has been 
used in~\cite{rrp} to formulate a criterion for identifying states
admitting LHV description.

Now, let $\hat\rho^{{\rm AB}}$ describe the state of a system of two
spin-1/2 particles, A and B. Following the approach outlined above,
the expression for the joint probability
$p(\epsilon^A_a,\epsilon^B_b)$ for the component $S_a$ of
spin A in the direction ${\bf a}$ to have value $\epsilon^A_a/2$, and
the component $S_b$ of spin B in direction ${\bf b}$ to have the value
$\epsilon^B_b/2$ ($\epsilon^A_a,\epsilon^B_b=\pm 1$) may be seen to be
given by,
\be
p(\epsilon^{A}_{a};\epsilon^{B}_{b})
={\rm Tr}\left[\left(\frac{1}{2}+\epsilon^{A}_{a}
\hat S^{\rm A}_a\right)\left(\frac{1}{2}
+\epsilon^{B}_{b}\hat S^{\rm B}_b\right)\hat\rho^{AB}\right].
\label{qd6}
\ee
In this case the issue of operator ordering does not arise as there
are no non-commuting operators in the product in the expression above.
The corresponding marginal distributions are
\be
p(\epsilon^{A}_{a})
&=&{\rm Tr}\left[\left(\frac{1}{2}+\epsilon^{A}_{a}
\hat S^{\rm A}_a\right)\hat\rho^{AB}\right]
\equiv \sum_{\epsilon^{\rm B}_b}p(\epsilon^{A}_{a};\epsilon^{B}_{b}),
\nonumber\\
p(\epsilon^{B}_{b})
&=&{\rm Tr}\left[\left(\frac{1}{2}+\epsilon^{B}_{b}
\hat S^{\rm B}_b\right)\hat\rho^{AB}\right]
\equiv \sum_{\epsilon^{\rm A}_a}p(\epsilon^{A}_{a};\epsilon^{B}_{b}).
\label{qd7n3}
\ee
The mutual information corresponding to $p(\epsilon^{A}_{a};
\epsilon^{B}_{b})$ in (\ref{qd6}) is
\be
I({\bf a},{\bf b})
=S(p(\epsilon^A_a))+S(p(\epsilon^B_b))
-S(p(\epsilon^A_a,\epsilon^B_b)),
\label{qd7n1}
\ee
where $S(p(\{x_i\}_n))$ is Shannon entropy for the probability 
$p(x_1,x_2,\cdots, x_n)\equiv p(\{x_i\}_n)$
of $n$ random variables $x_1,x_2,\ldots,x_n$:
\be
S((p(\{x_i\}_n))=-\sum_{\{x_i\}_n}p(\{x_i\}_n){\rm log}~p(\{x_i\}_n),
\label{qd7}
\ee
and the logarithm is to the base 2. The Eq.(\ref{qd7n1}) gives
the LHV theoretic classical expression for information
in the distribution function of the components in directions
${\bf a}$ and ${\bf b}$.

On the other hand, the quantum theoretic mutual information for the
system described by the density operator $\hat\rho^{{\rm AB}}$ is
given by
\be
I_{{\rm Q}}(\hat\rho^{{\rm AB}})=S_{\rm Q}(\hat\rho^{{\rm A}})+
S_{\rm Q}(\hat\rho^{{\rm B}})-S_{\rm Q}(\hat\rho^{{\rm AB}}),
\label{qd7n4}
\ee
where $S_{\rm Q}(\hat\rho)$ denotes the von Neumann entropy:
\be
S_{\rm Q}(\hat\rho)=-{\rm Tr}[\hat\rho{\rm log}(\hat\rho)],
\label{qd2}
\ee
and $\hat\rho^{{\rm A}}={\rm Tr}_{{\rm B}}(\hat\rho^{{\rm AB}})$,
$\hat\rho^{{\rm B}}={\rm Tr}_{{\rm A}}(\hat\rho^{{\rm AB}})$ are the
reduced density operators of the spins A and B respectively.

The LHV theoretic quantumness of mutual information for the joint
probability for the component of A in the direction ${\bf a}$ and that 
of B in the direction ${\bf b}$ to have the values $\pm1/2$ may be 
defined as
\be
Q({\bf a},{\bf b})&=&
I_{{\rm Q}}(\hat\rho^{{\rm AB}})
-I({\bf a},{\bf b}).
\label{qd8}
\ee
Different measures of quantumness are obtained by different choices of
the directions ${\bf a}$ and ${\bf b}$. It is proposed to specify 
${\bf a}$ and  ${\bf b}$ by noting the following:
\begin{enumerate}
\item The variance in the measurement of $\hat{\bf S}\cdot{\bf b}$ 
i.e. in the component of spin along the direction ${\bf b}$ is given by
\be
\Delta(\hat{\bf S}\cdot{\bf b})^2=\lang(\hat{\bf S}\cdot{\bf b})^2\rang
-\lang\hat{\bf S}\cdot{\bf b}\rang^2=\frac{1}{4}
-(\lang\hat{\bf S}\rang\cdot{\bf b})^2.
\label{dm4}
\ee
This shows that the variance is minimum when ${\bf b}$ is in the 
direction of $\lang\hat{\bf S}\rang$ i.e. in the direction of the
Bloch vector of the spin.

\item Let $|\pm,{\bf a}\rang$ denote the eigenstates of the spin
component $\hat{\bf S}\cdot{\bf a}$ in the direction ${\bf a}$.
Let $\hat\rho$ be the density matrix describibg the state of the
spin-1/2 particle and let $\hat\rho_a$ denote its state after 
measurement of its component along the direction ${\bf a}$. It can be 
shown that $S(\hat\rho)\le S(\hat\rho_a)$ with equality holding if and 
only if ${\bf a}$ is such that $|\pm,{\bf a}\rang$ is the eigenstates 
of $\hat\rho$~\cite{luo1}. Thus the least disturbing measurement is
along the direction ${\bf a}$ which is such that $|\pm,{\bf a}\rang$ 
are the eigenstates of $\hat\rho$. Now, recall that the density matrix
of a spin-1/2 particle may be written as
\be
\hat\rho=\frac{I}{2}+2\lang\hat{\bf S}\rang\cdot\hat{\bf S}.
\label{dm3}
\ee
This shows that the eigenstates of the spin component along the 
direction $\lang\hat{\bf S}\rang$ are the eigenstates of $\hat\rho$ as 
well. Hence the least disturbing measurement in the sense described 
above is along the direction of $\lang\hat{\bf S}\rang$.
\end{enumerate}
We thus see that the direction of $\lang\hat{\bf S}\rang$ i.e. the
direction of the Bloch vector has special significance as the one in
which the spin component has minimum variance and also the one
along which the measurement is least disturbing.
Note also that the criterion for 
identifying quantumness in the
correlations between observables in~\cite{rrp} is based on
the properties of the joint quasiprobability in symmetric ordering for
the eigenvalues of the components of each spin in three mutually
orthogonal directions, one of which is the direction of the Bloch 
vector of that spin, and on the said quasiprobability for two of the 
three  said components. That criterion identifies non-classical states 
of two or more spin-1/2 particles in agreement with the predictions 
based on other approaches, including the prediction of classicality of 
certain non-separable states. 

In view of the discussion above, we let ${\bf a}$ and ${\bf b}$ in 
(\ref{qd8}) to be the directions of the Bloch vectors of spins A and B 
respectively if those vectors are non-zero and define the quantumness 
of mutual information as~\cite{rrp2}
\be
Q_{{\rm LHV}}=I_{{\rm Q}}(\hat\rho^{{\rm AB}})-I_{{\rm LHV}},
\label{qd9p1}
\ee
where
\be
I_{{\rm LHV}}\equiv I({\bf a},{\bf b}),\quad
{\bf a}=\frac{\langle\hat{{\bf S}}^{\rm A}\rangle}
{|\langle\hat{{\bf S}}^{\rm A}\rangle|}\ne 0,\quad 
{\bf b}=\frac{\langle\hat{{\bf S}}^{\rm B}\rangle}
{|\langle\hat{{\bf S}}^{\rm B}\rangle|}\ne 0.
\label{nd1}
\ee
We will see that $Q_{{\rm LHV}}$ in this case is identical with the 
measurement induced disturbance. However, the measurement induced 
disturbance does not specify ${\bf a}$ (${\bf b}$) uniquely when
$\langle\hat{{\bf S}}^{\rm A}\rangle=0$
($\langle\hat{{\bf S}}^{\rm B}\rangle=0$) but, as discussed below, 
$I_{{\rm LHV}}$ can be specified and $Q_{{\rm LHV}}$ determined 
uniquely even in such cases.

\begin{enumerate}
\item Let $\langle\hat{{\bf S}}^{\rm A}\rangle=0$ but
$\langle\hat{{\bf S}}^{\rm B}\rangle\ne 0$. The
${\bf a}$ in (\ref{nd1}) can then be any direction. In that case
$I_{{\rm LHV}}$ is defined to be the maximum of
$I({\bf a},{\bf b})$ over all ${\bf a}$:
\be
I_{{\rm LHV}}={\rm max}_{{\bf a}}
I({\bf a},{\bf b}),\quad
\langle\hat{{\bf S}}^{\rm A}\rangle=0,\quad
{\bf b}=\frac{\langle\hat{{\bf S}}^{\rm B}\rangle}
{|\langle\hat{{\bf S}}^{\rm B}\rangle|}\ne 0.
\label{qd9pq1}
\ee
We will see that $Q_{{\rm LHV}}$ in this case is the same as 
quantum discord if $\hat\rho^{{\rm AB}}$ satisfies the condition 
specified following Eq.(\ref{qdd16}).

\item If $\langle\hat{{\bf S}}^{\rm A}\rangle=
\langle\hat{{\bf S}}^{\rm B}\rangle=0$, i.e.
$\langle\hat S^{\rm A}_a\rangle=\langle\hat S^{\rm B}_b\rangle=0$
for all ${\bf a}$ and ${\bf b}$ 
then both, ${\bf a}$ and ${\bf b}$, in (\ref{nd1}) are arbitrary.
In this case $I_{{\rm LHV}}$ is defined to be the maximum of 
$I({\bf a},{\bf b})$ over all directions ${\bf a}$ and ${\bf b}$:
\be
I_{{\rm LHV}}={\rm max}_{{\bf a},{\bf b}}
I({\bf a},{\bf b}),\quad
\langle\hat{{\bf S}}^{\rm A}\rangle=
\langle\hat{{\bf S}}^{\rm B}\rangle=0.
\label{qd9pq2}
\ee
By evaluating the expression above analytically exactly in the 
following, we will show that $Q_{{\rm LHV}}$ in this case is the same 
as the symmetric discord.

\vskip .1 in\noi
To that end, let $({\bf e}_1,{\bf e}_2,{\bf e}_3)$
be an orthonormal set of cartesian vectors and let 
${\bf a}=\sum_iu_i{\bf e}_i$, ${\bf b}=\sum_iv_i{\bf e}_i$
$\hat{{\bf S}}^\mu=\sum_i{\bf e}_i\hat S^\mu_i$ 
($\mu={\rm A, B}$) so that
\be
\hat S^{{\rm A}}_a={\bf a}\cdot\hat{\bf S}^{\rm A}
=\sum_{i=1}^{3}u_i\hat S^{\rm A}_i,\qquad
\hat S^{{\rm B}}_b={\bf a}\cdot\hat{\bf S}^{\rm B}
=\sum_{i=1}^{3}v_i\hat S^{\rm B}_i,
\label{qd25a}
\ee
and
\be
\langle\hat S^{{\rm A}}_a\hat S^{{\rm B}}_b\rangle
=\frac{1}{4}\sum_{i,j=1}^{3}C_{ij}u_iv_j,
\label{qd25b}
\ee
where
\be
{\rm C}_{ij}=4{\rm Tr}\left(\hat S^{\rm A}_i
\hat S^{\rm B}_j\hat\rho^{AB}\right),
\label{qd6nn3}
\ee
denotes correlation between the components of the spins A and B. The
expression (\ref{qd6}) for $p(\epsilon^{A}_{a};\epsilon^{B}_{b})$
then assumes the form
\be
p(\epsilon^{A}_{a};\epsilon^{B}_{b})
&=&{\rm Tr}\left[\left(\frac{1}{4}+\epsilon^{A}_{a}
\epsilon^{B}_{b}
\hat S^{\rm A}_{a}\hat S^{\rm B}_{b}\right)
\hat\rho^{AB}\right]\nonumber\\
&=&\frac{1}{4}\left(1+\epsilon^{A}_{a}\epsilon^{B}_{b}
\sum_{i,j=1}^{3}C_{ij}u_iv_j\right),
\label{qd6nn1}
\ee
and $p(\epsilon^{A}_a)=p(\epsilon^{B}_b)=1/2$. It can be shown that in
this case~\cite{wu}
\be
{\rm max}_{{\bf a},{\bf b}}
I({\bf a},{\bf b})=1+H\left(\frac{1+{\rm C}}{2},
\frac{1-{\rm C}}{2}\right),
\label{qd6nn2p2}
\ee
with
\be
H\left(\{x_i\}_N\right)=-\sum_{i=1}^{N}x_i{\rm log}(x_i),
\label{qd6nn2b}
\ee
and
\be
C={\rm max}(|f_1|,|f_2|,|f_3|),
\label{qd6nn2b}
\ee
where $(f_1,f_2,f_3)$ are the singular values of the matrix
$\hat C\equiv \{C_{ij}\}$. It then follows that
\be
Q_{{\rm LHV}}=I_{{\rm Q}}(\hat\rho^{{\rm AB}})
-\left[1+H\left(\frac{1+{\rm C}}{2},\frac{1-{\rm C}}{2}\right)\right].
\label{qd6nn2p}
\ee
This is an exact analytic expression for the LHV theoretic quantumness 
of mutual information when the Bloch vector of each spin vanishes.

\end{enumerate}

\section{Comparison with Other Measures}~\label{sec3}

Let us now compare the measure introduced above with other measures.
To that end, let $\prod^{\rm A}_{\pm\bf a}$ and
$\prod^{\rm B}_{\pm\bf b}$ be the complete set of one-dimensional
orthogonal projection operators for projective measurements in
directions ${\bf a}$ and ${\bf b}$ on spins A and B. The state of the
system after the measurement would be~\cite{luo1}
\be
\hat{\tilde\rho}^{{\rm AB}}({\bf a},{\bf b})=\sum_{\mu, \nu=\pm 1}
\hat{\prod}^{{\rm A}}_{\mu{\bf a}}
\otimes\hat{\prod}^{{\rm B}}_{\nu{\bf b}}\hat\rho^{{\rm AB}}
\hat{\prod}^{{\rm A}}_{\mu{\bf a}}\otimes
\hat{\prod}^{{\rm B}}_{\nu{\bf b}}.
\label{qd7n3}
\ee
By noting that
\be
\hat{\prod}^{{\rm A}}_{\pm{\bf a}}=\frac{1}{2}\pm\hat S^{{\rm A}}_{a},
\qquad
\hat{\prod}^{{\rm B}}_{\pm{\bf b}}=\frac{1}{2}\pm\hat S^{{\rm B}}_{b},
\label{qd8n2}
\ee
it is straightforward to see that
\be
I_{\rm Q}(\hat{\tilde\rho}^{{\rm AB}}({\bf a},{\bf b}))
=I({\bf a},{\bf b}).
\label{qd8n3}
\ee
In the following we use the results above for comparing the LHV 
theoretic and other measures of quantumness of mutual information.

\begin{enumerate}
\item Consider the measuremement induced disturbance measure defined
as
\be
Q_{{\rm MID}}=I_Q(\hat\rho^{{\rm AB}})
-I_Q(\hat{\tilde\rho}^{{\rm AB}}({\bf a},{\bf b})),
\label{qd8n4}
\ee
where $\prod^{\rm A}_{\pm\bf a}$ and $\prod^{\rm B}_{\pm\bf b}$ are
projections on the eigenbasis of the reduced density operators
$\hat\rho^{\rm A}$ and $\hat\rho^{\rm B}$ of spins A and B
respectively. A reason for the choice of measurement induced by 
projectors on the eigenbases of the density operators is that such 
measurements are least disturbing~\cite{luo1}. As shown
following Eq.(\ref{dm3}), in this case ${\bf a}$ and ${\bf b}$ are 
also the directions of $\lang\hat{{\bf S}}^{\rm A}\rang$ and 
$\lang\hat{{\bf S}}^{\rm B}\rang$. Thus, $\prod_{\pm a}$ are 
projectors on the eigenbasis of $\hat\rho^{\rm A}$ and ${\bf a}$ is 
also the direction of the Bloch vector of spin A with similar 
observation about the  spin B. From (\ref{qd8n3}) and (\ref{nd1}) it 
then follows that
\be
I_Q(\hat{\tilde\rho}({\bf a},{\bf b}))=I_{\rm LHV},\quad
{\bf a}=\frac{\langle\hat{{\bf S}}^{\rm A}\rangle}
{|\langle\hat{{\bf S}}^{\rm A}\rangle|}\ne 0,\quad 
{\bf b}=\frac{\langle\hat{{\bf S}}^{\rm B}\rangle}
{|\langle\hat{{\bf S}}^{\rm B}\rangle|}\ne 0.
\label{ndf3}
\ee
Hence, the measurement induced disturbance and LHV theoretic measures
are same when the Bloch vectors of the two spins are non-zero: 
\be
Q_{{\rm LHV}}=Q_{{\rm MID}},\quad
{\bf a}=\frac{\langle\hat{{\bf S}}^{\rm A}\rangle}
{|\langle\hat{{\bf S}}^{\rm A}\rangle|}\ne 0,\quad
{\bf b}=\frac{\langle\hat{{\bf S}}^{\rm B}\rangle}
{|\langle\hat{{\bf S}}^{\rm B}\rangle|}\ne 0.
\label{qd7n5}
\ee
In case $\langle\hat{{\bf S}}^{{\rm A}}\rangle=0$, (\ref{dm3})
shows that $\hat\rho^{{\rm A}}=I/2$ which means that the eigenbasis
of reduced density operator of A is not unique. In such cases, 
measurement induced disturbance is
not unique whereas $Q_{{\rm LHV}}$, evaluated as in (\ref{qd9pq1}),
is uniquely determined and will be shown below to be analogous to
quantum discord if $\hat\rho^{{\rm AB}}$ satisfies the condition 
specified following Eq.(\ref{qdd16}).

\item The quantum discord for projective measurement on A is defined
as~\cite{streltsov}
\be
Q_{{\rm D}}=S_{\rm Q}\left(\hat\rho^{\rm A}\right)
-S_{\rm Q}\left(\hat\rho^{\rm AB}\right)+\mbox{min}_{{\bf a}}
\sum_{\mu=\pm}p_\mu S_{\rm Q}\left(\hat\rho_\mu\right),
\label{qdd1}
\ee
where
\be
p_\mu={\rm Tr}\hat{\prod}^{{\rm A}}_{\mu{\bf a}}\hat\rho^{\rm AB}
\hat{\prod}^{{\rm A}}_{\mu{\bf a}},\qquad
\hat\rho_\mu=\frac{\hat{\prod}^{{\rm A}}_{\mu{\bf a}}\hat\rho^{\rm AB}
\hat{\prod}^{{\rm A}}_{\mu{\bf a}}}{p_\mu}.
\label{qdd2}
\ee
We have
\be
\sum_{\mu}p_\mu S_{\rm Q}\left(\hat\rho_\mu\right)
&=&S_{\rm Q}\left(\hat\rho^{\prime{\rm AB}}({\bf a})\right)
-S_{\rm Q}\left(\hat\rho^{\prime{\rm A}}({\bf a})\right)\nonumber\\
&=&S_{\rm Q}\left(\hat\rho^{{\rm B}}_{++}({\bf a})\right)+
S_{\rm Q}\left(\hat\rho^{{\rm B}}_{--}({\bf a})\right)
-S_{\rm Q}\left(\hat\rho^{\prime{\rm A}}({\bf a})\right),
\label{qdd5}
\ee
where 
\be
\hat\rho^{\prime{\rm AB}}({\bf a})
=\sum_{\mu=\pm}\hat{\prod}^{{\rm A}}_{\mu{\bf a}}
\hat\rho^{\rm AB}\hat{\prod}^{{\rm A}}_{\mu{\bf a}},\qquad
\hat\rho^{\prime{\rm A}}({\bf a})=
{\rm Tr}_{{\rm B}}(\hat\rho^{\prime{\rm AB}}({\bf a})),
\label{qdd4}
\ee
and
\be
\hat\rho^{{\rm B}}_{\mu\mu}({\bf a})=
\langle\mu{\bf a}|\hat\rho^{\rm AB}|\mu{\bf a}\rangle,\qquad \mu=\pm.
\label{qdd6}
\ee
Now, let optimization in (\ref{qdd1}) be attained for
${\bf a}={\bf a}_m$ then, on invoking (\ref{qdd5}), the last term in
(\ref{qdd1}) assumes the form
\be
\mbox{min}_{{\bf a}}\sum_{\mu=\pm}p_\mu S_{\rm Q}
\left(\hat\rho_\mu\right)
&=&S_{\rm Q}\left(\hat\rho^{{\rm B}}_{++}({\bf a}_m)\right)+
S_{\rm Q}\left(\hat\rho^{{\rm B}}_{--}({\bf a}_m)\right)\nonumber\\
&&-S_{\rm Q}\left(\hat\rho^{\prime{\rm A}}({\bf a}_m)\right).
\label{qddpq}
\ee
If ${\bf a}_m$ is such that the eigenbasis of
$\hat\rho^{{\rm B}}_{++}({\bf a}_m)$ and that of
$\hat\rho^{{\rm B}}_{--}({\bf a}_m)$ is the same as the eigenbasis
$|\pm{\bf b}\rangle$ of $\hat\rho^{{\rm B}}$ then
\be
&&S_{\rm Q}(\hat\rho^{\rm B}_{++}({\bf a}_m))
+S_{\rm Q}(\hat\rho^{\rm B}_{--}({\bf a}_m))\nonumber\\
&&=H\left(\langle {\bf b}|\hat\rho^{\rm B}_{++}|{\bf b}\rangle,
\langle -{\bf b}|\hat\rho^{\rm B}_{++}|-{\bf b}\rangle\right)
\nonumber\\
&&~~+H\left(\langle {\bf b}|\hat\rho^{\rm B}_{--}|{\bf b}\rangle,
\langle -{\bf b}|\hat\rho^{\rm B}_{--}|-{\bf b}\rangle\right)
\nonumber\\
&&=H\left(\langle {\bf a}_m,{\bf b}|\hat\rho^{\rm AB}
|{\bf a}_m,{\bf b}\rangle, \langle {\bf a}_m,-{\bf b}
|\hat\rho^{\rm AB}|{\bf a}_m,-{\bf b}\rangle\right)
\nonumber\\
&&~~+H\left(\langle -{\bf a}_m,{\bf b}|\hat\rho^{\rm AB}|
-{\bf a}_m,{\bf b}\rangle,
\langle -{\bf a}_m,-{\bf b}|\hat\rho^{\rm AB}
|-{\bf a}_m,-{\bf b}\rangle\right)\nonumber\\
&&=S\left(p(\epsilon^{\rm A}_{a_m},\epsilon^{\rm B}_b)\right),
\label{qdd6nn3}
\ee
where $p(\epsilon^{\rm A}_{a_m},\epsilon^{B}_b)$ is as in (\ref{qd6}).
Also, if $\langle\hat{\bf S}^{\rm A}\rangle=0$ then
$\hat\rho^{\rm A}=\hat\rho^{\prime{\rm A}}=I$. The expression 
(\ref{qdd1}) for the quantum discord may then be rewritten as
\be
Q_{\rm D}=I_Q(\hat\rho^{{\rm AB}})-I({\bf a}_m,{\bf b}).
\label{f1}
\ee
This may be interpreted as
\be
Q_{\rm D}=I_Q(\hat\rho^{{\rm AB}})
-\mbox{max}_{\bf a}I({\bf a},{\bf b}).
\label{f2}
\ee
By virtue of (\ref{qd9pq1}), the right hand side of this expression
is $Q_{\rm LHV}$. Thus we find that
\be
Q_{\rm LHV}=Q_{{\rm D}},\quad
\langle\hat{\bf S}^{\rm A}\rangle=0,\quad
\langle\hat{\bf S}^{\rm B}\rangle\ne 0.
\label{qdd16}
\ee
It should be emphasized that the result above is valid only when the
eigenbases of $\hat\rho^{\rm B}_{\pm,\pm}({\bf a}_m)$ in the optimum 
measurement are same as the eigenbasis of $\hat\rho^{\rm B}$.

\item Next, recall that the symmetric discord $Q_{{\rm SYM}}$ is
defined by
\be
Q_{{\rm SYM}}=I_{\rm Q}\left(\hat\rho^{\rm AB}\right)
-{\rm max}_{{\bf a},{\bf b}}I_{\rm Q}
\left(\hat{\tilde\rho}^{\rm AB}\right),
\label{ndef4}
\ee
where $\hat{\tilde\rho}^{\rm AB}$ is as in (\ref{qd7n3}). By recalling
(\ref{qd8n3}) it follows that
\be
Q_{{\rm SYM}}=I_{\rm Q}\left(\hat\rho^{\rm AB}\right)
-{\rm max}_{{\bf a},{\bf b}}I({\bf a},{\bf b}),
\label{ndef5}
\ee
which, on invoking (\ref{qd9pq2}), yields
\be
Q_{{\rm LHV}}=Q_{{\rm SYM}}, \quad
\langle\hat{{\bf S}}^{\rm A}\rangle=
\langle\hat{{\bf S}}^{\rm B}\rangle=0.
\label{ndef6}
\ee
Thus $Q_{{\rm SYM}}$ is same as $Q_{{\rm LHV}}$ if the Bloch vector
of each spins is zero.
\end{enumerate}

As examples, consider first the pure state. The Bloch vector of 
each spin in this case is non-zero. By virtue of the considerations 
above, it follows that $Q_{{\rm LHV}}$ then is same as the
measurement induced disturbance. It turns out to be the same also
as the symmetric and the quantum discords.

As regards mixed states, recall that any mixed state of two qubits can
be expressed as~\cite{luo2}
\be
\hat\rho^{\rm AB}=\frac{1}{4}I^{\rm A}\otimes I^{\rm B}
+\sum_{i,j=1,2,3}w_{ij}\hat S^{{\rm A}}_i\otimes\hat S^{{\rm B}}_j
+\frac{r}{2}\hat S^{\rm A}_3\otimes I^{\rm B}
+\frac{s}{2}I^{\rm A}\otimes\hat S^{\rm B}_3.
\label{qd15}
\ee
Here $\hat S^\mu_i={\bf e}_i\cdot\hat{{\bf S}}^\mu$ with $\mu=A,B$ and
${\bf e}_i\cdot{\bf e}_j=\delta_{ij}$. 
The optimization involved in symmetric and other discords is generally
a formidable task. We consider some special cases. 

\begin{enumerate}
\item Consider the case $r=s=0$. The expression (\ref{qd15}) then reads
\be
\hat\rho^{\rm AB}=\frac{1}{4}I^{\rm A}\otimes I^{\rm B}
+\sum_{i,j=1,2,3}w_{ij}\hat S^{{\rm A}}_i\otimes\hat S^{{\rm B}}_j,
\label{qd15n}
\ee
so that
\be
\langle\hat{{\bf S}^{\rm A}}\rangle
=\langle\hat{{\bf S}^{\rm B}}\rangle=0,\qquad
\langle\hat S^{A}_i\hat S^{B}_j\rangle&=&\frac{w_{ij}}{4}.
\label{qd16}
\ee
The exact expression for $Q_{\rm LHV}$ in this case is given by
(\ref{qd6nn2p}). Insert in it the expression for
$I_Q(\hat\rho^{\rm AB})$ with $\hat\rho^{\rm AB}$ given by 
(\ref{qd15n}). It can be shown that
\be
I_Q(\hat\rho^{{\rm AB}})=
2-H\left(\lambda_1,\lambda_2,\lambda_3,\lambda_4\right),
\label{ex1}
\ee
where $\lambda_i$ ($i=1,2,3,4$) are the eigenvalues of 
$\hat\rho^{{\rm AB}}$ given by
\be
\lambda_1&=&\frac{1-f_1-f_2-f_3}{4},\quad
\lambda_2=\frac{1-f_1+f_2+f_3}{4},\nonumber\\
\lambda_3&=&\frac{1+f_1-f_2+f_3}{4},\quad
\lambda_4=\frac{1+f_1+f_2-f_3}{4},
\label{ex2}
\ee
with $f_1,f_2,f_3$ being the eigenvalues of the $3\times 3$ matrix formed by
$w_{ij}$ ($i,j=1,2,3$) as its elements. Hence
\be
Q_{{\rm LHV}}=1-H\left(\lambda_1,\lambda_2,\lambda_3,\lambda_4\right)
+H\left(\frac{1+{\rm C}}{2},\frac{1-{\rm C}}{2}\right),
\label{ex3p}
\ee
where $C$ is the maximum of $(|f_1|,|f_2|,|f_3|)$.
In accordance with the assertion in Eq.(\ref{ndef6}), this is the same
as the expression for symmetric discord derived in~\cite{wu}.

Several states, like Werner state, and others for which analytic
results for various discords are available, fall in the
category of vanishing average directions of both the spins~\cite{wu}.
The $Q_{{\rm LHV}}$ for such states correspond to special cases of
(\ref{ex3p}).

\item Next, consider the following form of (\ref{qd15}) for which
analytic results for quantum discord are
known~\cite{ali}-\cite{sarandy},
\be
\hat\rho^{\rm AB}&=&\frac{1}{4}I^{\rm A}\otimes I^{\rm B}
+\sum_{i=1}^{3}c_{i}\hat S^{{\rm A}}_i\otimes\hat S^{{\rm B}}_i
+\frac{r}{2}\hat S^{{\rm A}}_3\otimes I^{\rm B}
+\frac{s}{2}I^{\rm A}\otimes \hat S^{{\rm B}}_3.
\label{ex3}
\ee
In this case
\be
\langle\hat{{\bf S}}^{\rm A}\rangle=\frac{r}{2}{{\bf e}_3}
=\langle\hat{{\bf S}}^{\rm B}\rangle=\frac{s}{2}{{\bf e}_3}.
\label{ex2-1}
\ee
Hence, if $r\ne 0$, $s\ne 0$, $I_{\rm LHV}=I({\bf a},{\bf b})$
with ${\bf a}={\bf b}={\bf e}_3$. The corresponding probabilities
are given by
\be
p(\epsilon^{\rm A},\epsilon^{\rm B})
=\frac{1}{4}\left[1+r\epsilon^{\rm A}+s\epsilon^{\rm B}
+c_3\epsilon^{\rm A}\epsilon^{\rm B}\right].
\label{ex2-2}
\ee
Using this, and the analytic expression for
$I_{\rm Q}(\hat\rho^{\rm AB})$~\cite{sarandy}, the LHV theoretic
quantumness $Q_{\rm LHV}$ can be easily evaluated. In accordance
with (\ref{qd7n5}), it is same as the measurement induced disturbance.

In case, say, $r=0$, the $I_{\rm LHV}$ is found by maximizing
$I({\bf a},{\bf e}_3)$ over all ${\bf a}$. It is found that ${\bf a}$
for which maxima is achieved corresponds to ${\bf a}={\bf e}_3$.
Hence, the appropriate LHV theoretic probability for computing
$Q_{\rm LHV}$ in this case is given by (\ref{ex2-2}) with $r=0$.
From the discussion circa (\ref{qdd16}), it follows that
$Q_{\rm LHV}$ will be same as quantum discord $Q_{\rm D}$ for
measurement over A if the direction of optimal measurement ${\bf a}_m$
is ${\bf e}_3$ and $|\pm{\bf e}_3\rangle$ is the eigenbasis
of $\hat\rho^{\rm B}_{\mu\mu}({\bf a}_m)$. In order to find the
conditions under which the said equality is achieved, we recall that,
assuming $|c_1|\ge |c_2|$, it has been shown in~\cite{ali}
that the direction ${\bf a}_m$ of optimal measurements on A for
quantum discord is ${\bf e}_3$ or ${\bf e}_1$. The results
of~\cite{ali} have been qualified in~\cite{chen} by deriving
conditions under which ${\bf a}_m$ is ${\bf e}_1$ or ${\bf e}_3$ or
none of the two. On invoking those results, the condition under
which ${\bf a}_m={\bf e}_3$ in the present case of $r=0$ reads
$c^2_1+r^2\le c^2_3$ and $|\pm{\bf e}_3\rangle$ constitutes eigenbasis
of $\hat\rho^{\rm B}_{\mu\mu}({\bf a}_m)$. Hence, under the said
condition, the LHV theoretic quantumness of mutual information and the
quantum discord are identical.

Lastly, the case of $r=s=0$ is the special case of (\ref{ex1})
corresponding to $w_{ij}=c_i\delta_{ij}$.
\end{enumerate}

\section{Conclusions}~\label{sec4}

By invoking explicitly the local hidden variable theory, a measure
of quantumness of mutual information $Q_{{\rm LHV}}$ for a system of 
two spin-1/2 particles is proposed. It is based on 
finding the difference between the quantum and classical mutual 
informations in which the classical mutual information 
corresponds to the joint probability of the eigenvalues of the 
spins each along a specified direction. 
The proposed measure circumvents the need of 
optimization when the Bloch vector of each spin is non-zero; the
optimization is needed but can be performed analytically exactly when
the Bloch vector of each spin vanishes and is simplified when the 
Bloch vector of only one of the spins is zero. In essence, the proposed 
measure is identical with the measurement 
induced disturbance when the Bloch vector of each of the spins is 
non-zero. However, whereas the measurement induced disturbance 
is non-unique when the Bloch vector of one or both the spins is zero, 
the proposed measure even then determines the quantumness of mutual
information unambiguously. The $Q_{{\rm LHV}}$ is identical with the 
symmetric discord if the Bloch vector of each spin vanishes. It is
same as the quantum discord if the Bloch vector of only one spin is 
zero and if the state in question possesses certain additional 
properties.

\end{document}